# Hopfield model with quasi-diagonal connection matrix

Leonid Litinskii (pensioner), Kiryat-Motzkin, Israel.

litinlb@gmail.com, tel: +972-55-937-4480, whatsapp: +7915-189-4759.

We analyze a Hopfield neural network with a quasi-diagonal connection matrix. We use the term "quasi-diagonal matrix" to denote a matrix with all elements equal zero except the elements on the first super- and sub-diagonals of the principle diagonal. The nonzero elements are arbitrary real numbers. Such matrix generalizes the well-known connection matrix of the one dimensional Ising model with open boundary conditions where all nonzero elements equal $+1$. We present a simple description of the fixed points of the Hopfield neural network and their dependence on the matrix elements. The obtained results also allow us to analyze the cases of a) the nonzero elements constitute arbitrary super- and sub-diagonals and b) periodic boundary conditions.

1. **Introduction**

We consider a symmetric matrix with nonzero elements only on its first super- and sub-diagonals:

$$\mathbf{A}_1 = \begin{pmatrix} 0 & a_{1,2} & 0 & 0 & \cdots & \cdots & 0 \\ a_{2,1} & 0 & a_{2,3} & 0 & \cdots & \cdots & 0 \\ 0 & a_{3,2} & 0 & a_{3,4} & 0 & \cdots & 0 \\ 0 & \cdots & \cdots & \cdots & \cdots & \cdots & 0 \\ 0 & \cdots & \cdots & a_{n-2,n-3} & 0 & a_{n-2,n-1} & 0 \\ 0 & \cdots & \cdots & 0 & a_{n-1,n-2} & 0 & a_{n-1,n} \\ 0 & 0 & \cdots & \cdots & 0 & a_{n,n-1} & 0 \end{pmatrix}. \qquad (1)$$

The matrices of such type appear frequently when we describe linear chains of binary "agents" with short-range interaction in the case of open boundary conditions. Then inside the chain each agent interacts with its nearest right and left neighbors only and the two agents at the ends of the chain interact with one (left or right) neighbor. When all the interaction constants $a_{i,i+1}$ are the same and $a_{i,i+1} \equiv 1$ the matrix (1) corresponds to the connection matrix of the one-dimensional Ising model.

Fixed points of the Hopfield neural network coincide with local minima of a quadratic binary functional [2]:

$$E(\mathbf{s}) = -(\mathbf{As},\mathbf{s}) = -\sum_{i,j=1}^{n} a_{i,j} s_i s_j, \text{ where } \mathbf{s} = (s_1, s_2, s_3,...,s_n), \ s_i = \pm 1, \ i=1,2,..,n.$$

Making use of physical definitions they usually refer to $E(\mathbf{s})$ as energy. At every step of the system evolution the value of $E(\mathbf{s})$ decreases and sooner or later the system finds itself in a local minimum. In this paper, we show that minima on the graph of the connection matrix coefficients $\{a_{i,i+1}\}_1^{n-1}$ as a function of the subscript $i+1$ define directly the minima of the energy $E(\mathbf{s})$. Since we need to use the word "*minimum*" when describing the graph of the coefficients $\{a_{i,i+1}\}_1^{n-1}$ as well as the minima of the energy $E(\mathbf{s})$, in what follows to avoid confusion we look for the *maxima* of the functional

$$F(\mathbf{s}) = -E(\mathbf{s}) = (\mathbf{As},\mathbf{s}) \to \max, \text{ where } \mathbf{s} = (s_1, s_2, s_3,...,s_n).$$

The maxima of this $F$-functional coincide with the energy minima and, consequently, with the fixed points of the neural network.

The same as in physics the binary variable whose values are equal 1 or $-1$ we call the spin and a configuration providing the global maximum of the functional $F(\mathbf{s})$ we call a ground state. Everywhere in what follows $\hat{s}_i$ denotes the value of the spin $s_i$.

Our direct calculations show (see Subsection 4.1) that a configuration $\hat{\mathbf{s}} = (\hat{s}_1, \hat{s}_2, \hat{s}_3, ..., \hat{s}_n)$ provides the global maximum of the $F$-functional when

$$\hat{s}_k = d_k \cdot \hat{s}_1, \text{ where } d_k = \left(\prod_{i=1}^{k-1} \varepsilon_{i,i+1}\right), \ k = 2, 3, ..., n; \quad (2)$$

the constants $\varepsilon_{i,i+1}$ are the signs of the nonzero matrix elements

$$\varepsilon_{i,i+1} = sign(a_{i,i+1}). \quad (3)$$

The value of the first coordinate may be $\hat{s}_1 = +1$ or $\hat{s}_1 = -1$ and the equalities (2) and (3) define all the remained coordinates.

According to equations (2) and (3) only the signs of the matrix elements $a_{i,i+1}$ define the ground state and the absolute values of $a_{i,i+1}$ define the maximal value of the $F$-functional

$$\max_{\mathbf{s}} F(\mathbf{s}) = F(\hat{\mathbf{s}}) = 2(|a_{12}| + |a_{23}| + ... + |a_{n-1n}|). \quad (4)$$

Choosing the signs of the matrix elements we obtain a connection matrix with an arbitrary ground state. For example, the ground state $\hat{s}_1 = \hat{s}_2 .. = \hat{s}_k = +1$, $\hat{s}_{k+1} = \hat{s}_{k+2} = .. = \hat{s}_n = -1$ corresponds to the connection matrix where only the term $a_{k,k+1}$ of the sequence $\{a_{i,i+1}\}_1^{n-1}$ is negative and all other terms are positive (see Eq. (2)). If not only matrix element $a_{k,k+1}$ is negative but the next matrix element is also negative, that is $a_{k+1,k+2} < 0$, the configuration

$$\hat{\mathbf{s}} = (+1, ..., +1, \underset{k+1}{-1}, +1 ... +1),$$

becomes the ground state. From the above arguments it follows that choosing the signs of the coefficients $a_{i,i+1}$ we can construct the ground state coinciding with any specified configuration.

In Section 2, we show that in the case of the connection matrix (1) with positive matrix elements $a_{i,i+1}$ it is not only possible to determine the ground state but to describe all the local maxima of the functional $F(\mathbf{s})$.

In Section 3, we extend the obtained results to connection matrices with only non-zero elements on the second sub- and super-diagonals or on the third sub- and super-diagonals, and so on.

In physics the connection matrix (1) corresponds to the Ising model with *open boundary conditions*. Namely, inside the chain each spin has two nearest neighbors and the spins $s_1$ and $s_n$ at the ends of the chain have only one nearest neighbor. The case of all spins interacting with their two nearest neighbors assumes the *periodic boundary conditions* that is we roll the chain of spins into a ring and introduce a nonzero interaction $a_{1n}$ between the spins $s_1$ and $s_n$. In Section 4, we generalize the results of Sections 2 and 3 to the cases of periodic boundary conditions as well as of alternating in signs matrix elements $a_{i,i+1}$.

In Conclusions, we discuss a direction of future development of our studies. The proof of the statements of Section 2 are in Appendix.

## 2. Maxima of $F$-functional with open boundary conditions. Generating configurations.

1) **Setting of problem**. Without loss of generality we can assume that the matrix elements $a_{i,i+1}$ are positive (see Subsection 4.1 point 1):

$$a_{i,i+1} > 0 \quad \forall i = 1, 2, .., n-1.$$

Then, the equations (2) - (4) are simplified and supposing $\hat{s}_1 = +1$ we obtain an obvious result: the configuration $\mathbf{s}_0 = (+1,+1,+1...,+1)$ provides the global maximum of the functional $F(\mathbf{s})$ and its value is

$$F_0 = F(\mathbf{s}_0) = 2(a_{12} + a_{23} + ... + a_{n-1n}). \tag{5}$$

A fundamental result of the associative neural network theory [4] states that for a configuration $\mathbf{s}$ to provide a maximum of the functional $F(\mathbf{s}) = (\mathbf{As},\mathbf{s})$ it is necessary and sufficient the coincidence of the signs of the coordinate $s_i$ and the local field $h_i = a_{i,i-1}s_{i-1} + a_{i,i+1}s_{i+1}$ for each value of the index $i$:

$$h_i s_i > 0, \quad i = 1, 2, ..., n. \tag{6}$$

When applying this result to the connection matrix (1), we obtain that to be a local maximum of the $F$-functional a configuration has to satisfy inequalities:

$$h_i s_i = \begin{cases} a_{12}s_2 s_1 > 0, & i = 1 & (7.1) \\ a_{i,i-1}s_{i-1}s_i + a_{i,i+1}s_{i+1}s_i > 0, & 2 < i < n-1 & (7.2) \\ a_{n,n-1}s_{n-1}s_n > 0, & i = n. & (7.3) \end{cases}$$

A simple formula

$$F(\mathbf{s}) = (\mathbf{As},\mathbf{s}) = \sum_{i=1}^{n} h_i s_i \tag{8}$$

relates the value of the $F$-functional with the local fields $h_i$.

It is clear that if all the constants $a_{i,i+1}$ are positive, the configuration $\mathbf{s}_0 = (+1,+1,+1...,+1)$ provides the global maximum of the functional $F(\mathbf{s})$ and with regard to Eq. (5) its value equals: $F(\mathbf{s}_0) = 2 \cdot \sum_{i=1}^{n-1} a_{i,i+1}$. However, even in this case there may be solutions of the system (7) with not all positive coordinates. In Appendix, we prove this not evident statement rigorously; here we only present our main arguments and the basic result.

To solve the set of inequalities (7) we begin with choosing the initial value of the first spin; for example let it be $\hat{s}_1 = +1$. Then step by step we obtain the values of all other spins from the condition of positivity of the right-hand sides of the inequalities (7). Suppose that we have found the optimal values $\hat{s}_{i-1} = +1$, $\hat{s}_i = +1$, and it is necessary to find the optimal value of the variable $s_{i+1}$. If in Eq. (7.2) $a_{i,i-1} < a_{i,i+1}$, the positive value of the sum in the right-hand side of this equation is possible only when the value of the variable $s_{i+1}$ is $\hat{s}_{i+1} = +1$. However, when $a_{i,i-1} > a_{i,i+1}$, there are two possibilities: $\hat{s}_{i+1} = \pm 1$. In this case, it is necessary to check, if we can use the both values of $s_{i+1}$ as initial to carry on the calculation for larger indices $i$ with account for the key condition (6) at each step. As a rule, we can do this for only one of the possible values $\hat{s}_{i+1}$, but in the most interesting cases the both initial values $\hat{s}_{i+1} = \pm 1$ allow us to proceed. This means that at $i+1$ the unique solution of the set of inequalities (7) splits into two co-equal solutions or, in other words, into positive and negative branches:

$$\hat{s}_{i+1}^{(+)} = \hat{s}_{i+2}^{(+)} = \hat{s}_{i+3}^{(+)} = ... = +1 \text{ and } \hat{s}_{i+1}^{(-)} = \hat{s}_{i+2}^{(-)} = \hat{s}_{i+3}^{(-)} = ... = -1.$$

In this Section we only formulate our main result; the proof is in Appendix. We would like to point out that the results described in Subsection 2.2 relates to the case of positive coefficients $a_{i,i+1}$. The alternating in sign coefficients we analyze in Section 4.

We remind that always we choose $\hat{s}_1 = +1$. In what follows for our analysis we use a schematic graph of the dependence of the positive constants $\{a_{i,i+1}\}_{i=1}^{n-1}$ on the value of the index $i+1$. As an example we consider the graph in Fig.1 with three minima *inside* the interval $i+1 \in [2,n]$. We would like to point out that the possible minima at the ends of the interval $[2,n]$ are not important.

## 2) Main theorem

*Let the graph of a sequence of the coefficients* $\{a_{i,i+1}\}_{i=1}^{n-1}$ *has inside the interval* $[2,n]$ $p$ *inner minima at the points* $l+1$, $t+1$, $u+1$,... *and* $2 < l+1 < t+1 < u+1 < ... < n$. *Then the* $F$ *-functional (8) has exactly* $2^p$ *maxima at the configurations with the coordinates obtained when constructing a binary "tree" of options with the aid of the algorithm* $1° - 5°$:

$1°$. *We set* $\hat{s}_1^{(+)} = +1$ *and assign the same value to the next spin variables* $s_i$ *with the indices up to* $i = l$ *where* $l+1$ *is the point of the first minimum of the sequence* $\{a_{i,i+1}\}_{i=1}^{n-1}$:

$$\hat{s}_1^{(+)} = \hat{s}_2^{(+)} = ... = \hat{s}_l^{(+)} = +1. \tag{9}$$

*The superscript* $(+)$ *indicates that we are at the first step of the configuration construction and all the coordinates are equal* $+1$.

$2°$. *At the point* $l+1$, *a splitting of the spin sequence into positive and negative branches occurs. Up to the point* $i = t$ *preceding the second minimum of the sequence* $\{a_{i,i+1}\}_{i=1}^{n-1}$, *we continue each new branch with an initially assigned positive or negative value of spin:*

$$\hat{s}_1^{(+)} = +1, \quad ... \quad \hat{s}_l^{(+)} = +1, \quad \nearrow \begin{array}{l} \hat{s}_{l+1}^{(++)} = +1, \quad ... \quad \hat{s}_t^{(++)} = +1 \\ \searrow \hat{s}_{l+1}^{(+-)} = -1, \quad ... \quad \hat{s}_t^{(+-)} = -1. \end{array}$$

*The upper superscript* $(++)$ *marks the elements of the positive branch obtained as the result of splitting of the sequence (9). The superscript* $(+-)$ *marks the elements of the negative branch.*

$3°$. *At the point* $t+1$, *each of these two new sequences splits into positive and negative branches again. Up to the point* $i = u$ *preceding the third minimum of the sequence* $\{a_{i,i+1}\}$, *we continue each of the four new branches with the initial values of the coordinates assigned at the point* $t+1$:

$$\hat{s}_1^{(+)} = +1, \quad ... \quad \hat{s}_l^{(+)} = +1, \begin{array}{l} \nearrow \hat{s}_{l+1}^{(++)} = +1, \quad ... \quad \hat{s}_t^{(++)} = +1, \begin{array}{l} \nearrow \hat{s}_{t+1}^{(+++)} = +1 \quad ...\hat{s}_u^{(+++)} = +1. \\ \searrow \hat{s}_{t+1}^{(++-)} = -1, \quad ...\hat{s}_u^{(++-)} = -1. \end{array} \\ \searrow \hat{s}_{l+1}^{(+-)} = -1, \quad ... \quad \hat{s}_t^{(+-)} = -1, \begin{array}{l} \nearrow \hat{s}_{t+1}^{(+-+)} = +1, \quad ...\hat{s}_u^{(+-+)} = +1. \\ \searrow \hat{s}_{t+1}^{(+--)} = -1, \quad ...\hat{s}_u^{(+--)} = -1. \end{array} \end{array}$$

*The upper superscripts* $(+++)$ *and* $(++-)$ *mark the elements of the positive and negative branches appeared due to splitting of the* $(++)$ *branch, respectively. Similarly, the superscripts* $(+-+)$ *and* $(+--)$ *mark the elements of the positive and negative branches of the solutions appeared as a result of splitting of the sequence* $(+-)$ *at the point* $t+1$.

$4°$. *Finally, at the point* $u+1$ *each of the four constructed sequences splits into positive and negative branches. Up to the last value of the index* $i+1 = n$ *we continue each of the new branches with the assigned initial spin value* $\hat{s}_{u+1}$ *at the point* $u+1$. *As a result we obtain* $2^3 = 8$ *configurations that are maxima of the* $F$ *-functional.*

$$
\begin{array}{l}
\phantom{}\nearrow \hat{s}_{u+1}^{(+++)} = +1, \ldots \hat{s}_{n}^{(+++)} = +1. \\
\phantom{x}\nearrow \hat{s}_{t+1}^{(++)} = +1, \ldots \hat{s}_{u}^{(+++)} = +1, \to \hat{s}_{u+1}^{(+++-)} = -1, \ldots \hat{s}_{n}^{(+++-)} = -1. \\
\phantom{xxxxxxxxx}\hat{s}_{l+1}^{(++)} = +1, \ldots \hat{s}_{t}^{(++)} = +1, \phantom{xxx}\nearrow \hat{s}_{u+1}^{(++-+)} = +1, \ldots \hat{s}_{n}^{(++-+)} = +1. \\
\phantom{xxxxx}\nearrow \phantom{x}\searrow \hat{s}_{t+1}^{(++-)} = -1, \phantom{x} \hat{s}_{u}^{(++-)} = -1, \to \hat{s}_{u+1}^{(++--)} = -1, \ldots \hat{s}_{n}^{(++--)} = -1. \\
\hat{s}_{1}^{(+)} = +1, \ldots \hat{s}_{l}^{(+)} = +1, \\
\phantom{xxxxx}\searrow \phantom{x}\nearrow \hat{s}_{t+1}^{(+-+)} = +1, \ldots \hat{s}_{u}^{(+-+)} = +1, \to \hat{s}_{u+1}^{(+-++)} = +1, \ldots \hat{s}_{n}^{(+-++)} = +1. \\
\phantom{xxxxxxxxx}\hat{s}_{l+1}^{(+-)} = -1, \phantom{x} \hat{s}_{t}^{(+-)} = -1, \phantom{xxx}\searrow \hat{s}_{u+1}^{(+-+-)} = -1, \ldots \hat{s}_{n}^{(+-+-)} = -1. \\
\phantom{x}\searrow \hat{s}_{t+1}^{(+--)} = -1, \ldots \hat{s}_{u}^{(+--)} = -1, \to \hat{s}_{u+1}^{(+--+)} = +1, \ldots \hat{s}_{n}^{(+--+)} = +1. \\
\phantom{}\searrow \hat{s}_{u+1}^{(+---)} = -1, \ldots \hat{s}_{n}^{(+---)} = -1.
\end{array}
$$

$5°$. *In the general case, when the sequence $\{a_{i,i+1}\}_{i=1}^{n-1}$ has $p$ a local minima inside the interval $i \in [2, n]$ we obtain $2^p$ $n$-dimensional configurations. Since at the each step the inequality (6) is satisfied the corresponding configurations are the maxima of the $F$-functional and this functional has no other maxima.*

We call the *max-configurations* the configurations where the $F$-functional achieves its maxima.

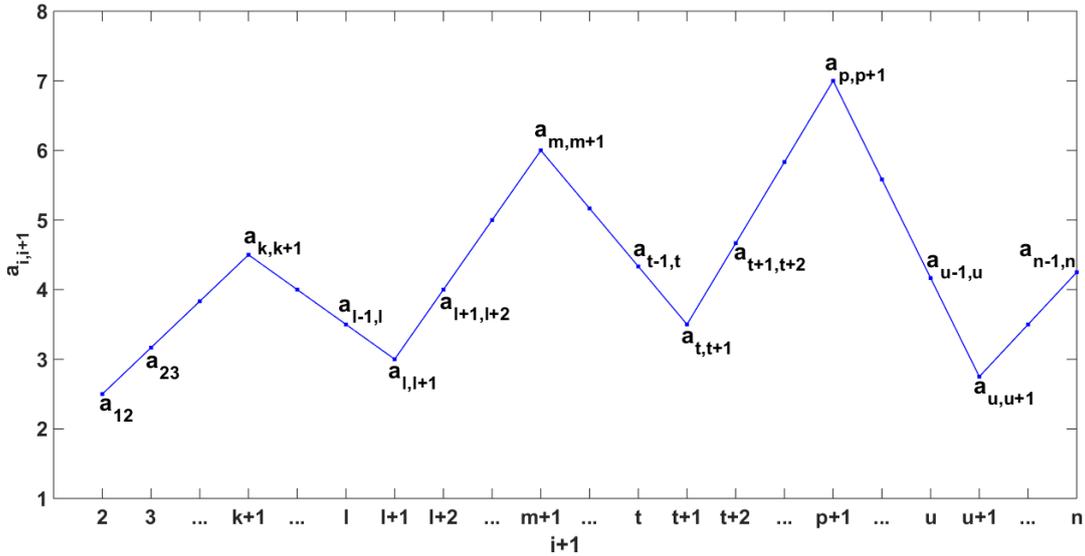

Fig. 1. Sketch of sequence $\{a_{i,i+1}\}_{i+1}^{n-1}$ with three inner minima at points $l+1$, $t+1$ and $u+1$.

Table 1. Max-configurations and maximal values of $F$ - functional

| | 1 | ... | $l$ | $l+1$ | ... | $t$ | $t+1$ | ... | $u$ | $u+1$ | ... | $n$ | |
|---|---|---|---|---|---|---|---|---|---|---|---|---|---|
| $\mathbf{s}^{(++++)}$ | +1 | ... | +1 | +1 | ... | +1 | +1 | ... | +1 | +1 | ... | +1 | $F_{++++} = F_0 = 2\sum_{i=1}^{n-1} a_{i,i+1}$ |
| $\mathbf{s}^{(+++-)}$ | +1 | ... | +1 | +1 | ... | +1 | +1 | ... | +1 | -1 | ... | -1 | $F_{+++-} = F_0 - 4a_{u,u+1}$ |
| $\mathbf{s}^{(++-+)}$ | +1 | ... | +1 | +1 | ... | +1 | -1 | ... | -1 | +1 | ... | +1 | $F_{++-+} = F_0 - 4(a_{t,t+1} + a_{u,u+1})$ |
| $\mathbf{s}^{(++--)}$ | +1 | ... | +1 | +1 | ... | +1 | -1 | ... | -1 | -1 | ... | -1 | $F_{++--} = F_0 - 4a_{t,t+1}$ |
| $\mathbf{s}^{(+-++)}$ | +1 | ... | +1 | -1 | ... | -1 | +1 | ... | +1 | +1 | ... | +1 | $F_{+-++} = F_0 - 4(a_{l,l+1} + a_{t,t+1})$ |
| $\mathbf{s}^{(+-+-)}$ | +1 | ... | +1 | -1 | ... | -1 | +1 | ... | +1 | -1 | ... | -1 | $F_{+-+-} = F_0 - 4(a_{l,l+1} + a_{t,t+1} + a_{u,u+1})$ |
| $\mathbf{s}^{(+--+)}$ | +1 | ... | +1 | -1 | ... | -1 | -1 | ... | -1 | +1 | ... | +1 | $F_{+--+} = F_0 - 4(a_{l,l+1} + a_{u,u+1})$ |
| $\mathbf{s}^{(+---)}$ | +1 | ... | +1 | -1 | ... | -1 | -1 | ... | -1 | -1 | ... | -1 | $F_{+---} = F_0 - 4a_{l,l+1}$ |

In Table 1, we show the explicit form of the obtained configurations for $p = 3$; in the last column are the values of the $F$-functional for each configuration.

The first configuration $\mathbf{s}^{(++++)}$ is the ground state of the $F$-functional. This configuration corresponds to the case when the spin variable equals $+1$ at all four intervals of the sign constancy. The configuration $\mathbf{s}^{(+++-)}$ corresponds to the case when the spin variables equal $+1$ everywhere except the last interval of the sign constancy where they are $-1$. In general, minus in the superscripts or subscripts denotes the interval of the spin variables equal $-1$.

Let us explain the values of the maxima of the $F$-functional in the last column of Table. In our system only the neighboring spins interact. With regard to Eq. (5) the value of the $F$-functional for the configuration $\mathbf{s}^{(++++)}$ equals $F_0$. To obtain the values of the $F$-functional for other configurations we take into account that if the two neighboring spins have the same signs their contribution to the $F$-functional is positive and in the case of the opposite signs, the contribution is negative. Consequently, when the spin signs change and the new interval of the sign constancy appears the interaction of spins at the ends intervals provides a negative contribution to the value of the $F$-functional. Compared to the global maximum $F_0$ the value of the corresponding local maxima decreases by four times the value of the energy of interaction of these two spins (see Eq. (A4) and Eq. (A5)).

**3) Generating configurations**

Let's call "generating" the three configurations from the first column of Table 1 characterized by one change of sign and the configuration of the global maximum:

$$\mathbf{s}^{(+++-)}, \quad \mathbf{s}^{(++--)}, \quad \mathbf{s}^{(+---)}, \quad \text{and} \quad \mathbf{s}^{(++++)}. \tag{10}$$

It turns out that with the aid of these configurations we can construct all other configurations providing maxima of the $F$-functional. We explain this statement constructing as example the configuration $\mathbf{s}^{(++-+)}$ where the sign of spins changes twice: from plus to minus at $t+1$ and from minus to plus at $u+1$.

Firstly, it is easy to see that we can express the value of $F_{++-+}$ with the aid of $F_{+++-}$, $F_{++--}$, and $F_{++++}$:

$$F_{++-+} = F_{+++-} + F_{++--} - F_{++++}, \tag{11}$$

Secondly, we can present the configuration $\mathbf{s}^{(++-+)}$ as a result of the coordinate-wise multiplication of the configurations we use in Eq. (11):

$$\mathbf{s}^{(++-+)} = \mathbf{s}^{(+++-)} \odot \mathbf{s}^{(++--)} \odot \mathbf{s}^{(++++)}; \tag{12}$$

the symbol $\odot$ denotes the coordinate-wise multiplication of vectors. Similar equations hold for other two max-configurations with two changes of signs: $\mathbf{s}^{(+-++)}$ and $\mathbf{s}^{(+--+)}$.

In the same time to describe the only one configuration $\mathbf{s}^{(+-+-)}$ whose coordinates change their signs three times we need all four generating configurations:

$$F_{+-+-} = F_{+++-} + F_{++--} + F_{+---} - 2F_{++++}, \quad \mathbf{s}^{(+-+-)} = \mathbf{s}^{(+++-)} \odot \mathbf{s}^{(++--)} \odot \mathbf{s}^{(+---)} \odot \mathbf{s}^{(++++)}. \tag{13}$$

When the coefficients $\{a_{i,i+1}\}_{i=1}^{n-1}$ have $p$ local minima the number of the generating configurations equals $p+1$. With the aid of equalities analogous to (11) – (13), we can express the characteristics of all other $2^p - (p+1)$ max-configurations using the generating configurations. The description of a large number of the max-configurations with the aid of a much less set of the generating configurations is important for analysis of the $F$-functional maxima. However, this simplification takes place only for the considered class of the matrices. Computer simulations show that in the general case there is no such a possibility.

**3. Second, third, … quasi-diagonals**

Let us consider a symmetric $n$-dimensional matrix with nonzero matrix elements only on the second sub- and super-diagonals:

$$\mathbf{A}_2 = \begin{pmatrix} 0 & 0 & a_{13} & 0 & \ddots & \ddots & \ddots & \vdots \\ 0 & 0 & 0 & a_{24} & 0 & \ddots & \ddots & \vdots \\ a_{31} & 0 & 0 & 0 & a_{35} & 0 & \ddots & \vdots \\ 0 & a_{42} & 0 & 0 & 0 & a_{46} & 0 & \vdots \\ \ddots & 0 & a_{53} & 0 & 0 & 0 & a_{57} & \vdots \\ \ddots & \ddots & 0 & a_{64} & 0 & 0 & 0 & \ddots \\ \ddots & \ddots & \ddots & 0 & a_{75} & 0 & 0 & \ddots \\ \cdots & \cdots & \cdots & \cdots & \cdots & \ddots & \ddots & \ddots \end{pmatrix}. \qquad (14)$$

For simplicity we assume that $n = 2m$.

It is easy to see that the set of inequalities (6) decomposes on two subsets: the first one describes $m$ spins with odd numbers and the other $m$ spins with even numbers are described by the second subset. Then in place of the inequalities (7) we have the necessary and sufficient conditions for a configuration to be maximum of the functional $F_2(\mathbf{s}) = (\mathbf{A}_2 \mathbf{s}, \mathbf{s})$ in the form:

$$h_i s_i = \begin{cases} a_{13} s_3 s_1 > 0, & i = 1 \\ (a_{31} s_1 + a_{53} s_5) s_3 > 0, & i = 3 \\ (a_{53} s_3 + a_{57} s_7) s_5 > 0, & i = 5 \\ \vdots & \vdots \\ (a_{n-3,n-5} s_{n-5} + a_{n-3,n-1} s_{n-1}) s_{n-3} > 0, & i = n-3 \\ a_{n-1,n-3} s_{n-3} s_{n-1} > 0, & i = n-1 \end{cases} \qquad (15)$$

and

$$h_i s_i = \begin{cases} a_{24} s_4 s_2 > 0, & i = 2 \\ (a_{42} s_2 + a_{46} s_6) s_4 > 0, & i = 4 \\ (a_{64} s_4 + a_{68} s_8) s_6 > 0, & i = 6 \\ \vdots & \vdots \\ (a_{n-2,n-4} s_{n-4} + a_{n-2,n} s_n) s_{n-2} > 0, & i = n-2 \\ a_{n,n-2} s_{n-2} s_n > 0, & i = n. \end{cases} \qquad (16)$$

The sets of the inequalities (15) and (16) describe two independent spin subsystems. Each subsystem of $m$ spins is similar to the system analyzed in Section 2. To find the maxima of the $F$-functional of the whole system (14) we proceed as follows. For each subsystem we plot the graphs of the corresponding coefficients $\{a_{2i-1,2i+1}\}_{i=1}^{m-1}$ and $\{a_{2i,2(i+1)}\}_{i=1}^{m-1}$. On each graph we determine the number and positions of the inner local minima and construct the Tables analogous to our Table. The last step is to combine the obtained results alternating numbers related to odd and even spins.

Suppose $p_1$ and $p_2$ are the numbers of the local minima on the first and the second graphs, respectively. Then the $F$-functional of the system (14) has $2 \cdot 2^{p_1 + p_2}$ maxima. The factor 2 appears because when we combine $m$-dimensional vectors $\mathbf{s}^{(1)}$ and $\mathbf{s}^{(2)}$ describing two different subsystems it is also necessary to account for the pair $(\mathbf{s}^{(1)}, -\mathbf{s}^{(2)})$ for which the value of the $F$-functional is the same as for the pair $(\mathbf{s}^{(1)}, \mathbf{s}^{(2)})$ but the corresponding $n$-dimensional configuration vectors differ. Consequently, in this case each local maximum of the $F$-functional is twice degenerate.

We can easily generalize the above-said to a connection matrix with nonzero elements only on the $k$-th quasi-diagonals. Let us describe in short the case $k = 3$. Now we have three independent spin subsystems three times smaller in dimension. For each subsystem we at first construct the set of the max-configurations of the smaller dimensions and after that combine them into $n$-dimensional configurations alternating data relating to different subsystems. Let $p_1, p_2$, and $p_3$ be the numbers of the inner local minima on the graphs corresponding to the three subsystems. Taking into account that each maxima of the $F$-functional is four time degenerate and consequently the number of the local maxima of the $F$-functional of the whole spin system equals $4 \cdot 2^{p_1+p_2+p_3}$.

We continue in the same manner when we analyze the quasi-diagonals with larger numbers $k$ and consequently with the larger number of the spin subsystems.

### 4. Periodic boundary conditions

In the preceding Sections we examined the spin chains supposing that unlike the spins inside the chain each of the end spins $s_1$ and $s_n$ have only one nearest neighbor and that all the matrix elements are positive. The same as in the Ising model we can introduce an interaction $a_{1n}$ between the end spins. The case when $a_{1n} = 0$ corresponds to *open boundary conditions*. If the chain resembles a ring and the interaction $a_{1n}$ is nonzero, they say that we deal with *periodic boundary conditions*.

In this section we generalize our results and discuss the cases of a) periodic boundary conditions and b) alternating in sign matrix elements

Bellow by a *spectrum* of the $F$-functional we understand the set of its maxima. We also recall that the term max-configuration denotes a configuration vector corresponding to the $F$-functional maximum.

**4.1. Global maximum of $F$-functional.** It is clear that in the case of periodic boundary conditions a cyclic replacement of all $n$ coefficients $a_{12}, a_{23}, ..., a_{n-1,n}, a_{1n}$ does not change the spectrum of the $F$-functional. Evidently, that when shifting cyclically the coefficients $a_{12}, a_{23}, ..., a_{n-1,n}, a_{1n}$ by $k$ positions to the left we obtain a new $F$-functional whose max-configurations are the result of cyclic shifting of the coordinates of the initial max-configurations by $k$ positions to the left. The same is also true in the case of open boundary conditions.

Let us obtain an expression for the $F$-functional global maximum in the case of the periodic boundary conditions supposing *different* signs of the matrix elements. This expression generalizes Eq. (4) corresponding to the open boundary conditions.

Without loss of generality we assume that the matrix element $a_{1n}$ is the smallest in modulus:

$$|a_{1n}| < |a_{i,i+1}| \ \forall \ i = 1,...,n-1 \ . \tag{17}$$

Our goal is to maximize the summand $h_i s_i$ for each value of $i$:

$$h_i s_i \xrightarrow{\max} \hat{h}_i \hat{s}_i = \left( a_{i,i-1} \hat{s}_{i-1} + a_{i,i+1} \hat{s}_{i+1} \right) \hat{s}_i \ .$$

We perform the maximization procedure step by step.

At the first step, we choose $\hat{s}_1 = +1$ and obtain the problem: $a_{12} s_2 + a_{1n} s_n \to \max$. We maximize the term $a_{12} s_2$ when $\hat{s}_2 = \varepsilon_{12}$ ($\varepsilon_{12}$ is defined by Eq. (2)). This choice guarantees that the sum of the both terms is positive for an arbitrary value of $s_n$. The optimal value $\hat{s}_n$ we define at the end of our analysis. Then

$$|a_{12}| + a_{1n} \hat{s}_n \to \max \ . \tag{18}$$

At the next step we have to maximize the term

$$a_{21}\hat{s}_1\hat{s}_2 + a_{23}s_3\hat{s}_2 = |a_{12}| + a_{23}\varepsilon_{12}s_3 \to \max .$$

To do this we set $\hat{s}_3 = \varepsilon_{12} \cdot \varepsilon_{23} = d_3$ (see Eq. 3). Then the maximal possible value of $h_2 s_2$ is $\hat{h}_2 \hat{s}_2 = |a_{12}| + |a_{23}|$.

We proceed in the same way until we find the optimal value $\hat{s}_{n-1}$ of the coordinate $s_{n-1}$:

$$\hat{s}_4 = \prod_{i=1}^{3}\varepsilon_{i,i+1} = d_4, \ \hat{s}_5 = \prod_{i=1}^{4}\varepsilon_{i,i+1} = d_5, ..., \hat{s}_{n-1} = \prod_{i=1}^{n-2}\varepsilon_{i,i+1} = d_{n-1}.$$

At each step, the choice of $\hat{s}_k$ maximizes the summand $h_{k-1}s_{k-1}$.

At the $(n-1)$-th step we have

$$a_{n-1,n-2}\hat{s}_{n-2}\hat{s}_{n-1} + a_{n-1,n}s_n\hat{s}_{n-1} = |a_{n-1,n-2}| + |a_{n-1,n}| \cdot s_n \cdot \prod_{i=1}^{n-1}\varepsilon_{i,i+1} . \qquad (19)$$

To maximize the expression (19), we set

$$\hat{s}_n = \prod_{i=1}^{n-1}\varepsilon_{i,i+1} = d_n . \qquad (20)$$

Finally, at the last $n$-th step the problem is

$$a_{n,n-1}\hat{s}_{n-1}\hat{s}_n + a_{n,1}\hat{s}_1\hat{s}_n = |a_{n,n-1}| + a_{1n}\hat{s}_n \to \max . \qquad (21)$$

Compared with Eq. (18) this equation does not contain any additional restrictions at $\hat{s}_n$ and the optimal value $\hat{s}_n$ we determined in Eq. (20). The question is whether $\varepsilon_{1n} = sign(a_{1n})$? If the signs of $\varepsilon_{1n}$ and $\hat{s}_n$ are the same the second terms in the expressions (18) and (21) are positive. If $\varepsilon_{1n}$ and $\hat{s}_n$ defined by Eq. (20) have different signs the second terms in these both expressions are negative. In any case the expressions (18) and (21) are positive since the modulus of the matrix element $|a_{1n}|$ is the smallest.

We conclude that

$$\max F(\mathbf{s}) = 2\left(\sum_{i=1}^{n-1}|a_{i,i+1}| + \varepsilon|a_{1n}|\right), \text{ where } \varepsilon = \varepsilon_{1n} \cdot \prod_{i=1}^{n-1}\varepsilon_{i,i+1} . \qquad (22)$$

In the case of the periodic boundary conditions the global maximum of the $F$-functional can be either more or less than the global maximum of the $F$-functional for the same system with open boundary conditions. The product of signs of $n$ elements of the connection matrix defines the result.

**4.2. Results of computer simulations.** In this Subsection we generalize the results of a large number of computer simulations.

**1). Relation between spectrums and max-configurations of $F(a)$- and $F(m)$-functionals.**

Let $F(a)$- and $F(m)$-functionals correspond to the alternating in sign matrix elements $a_{i,i+1}$ and the matrix elements equal their modulus $|a_{i,i+1}|$, respectively. By $\mathbf{s}^{(k)} = \left(s_1^{(k)}, s_2^{(k)}, ..., s_n^{(k)}\right)$, $k = 1, 2, ..., q_a$ and $\boldsymbol{\sigma}^{(l)} = \left(\sigma_1^{(l)}, \sigma_2^{(l)}, ..., \sigma_n^{(l)}\right)$, $l = 1, 2, ..., q_m$ we denote the sets of the max-configurations of $F(a)$- and $F(m)$-functionals, respectively. We discuss open and periodic boundary conditions separately.

*Open boundary conditions.* In this case, the spectrums of $F(a)$- and $F(m)$-functionals coincide: $F^{(k)}(a) = F^{(k)}(m)$, $k = 1, 2, ..., q = q_a = q_m$. The coordinates of the max-configurations of these functionals are related with the aid of the known coefficients $d_k$ defined in Eq. (3):

$$\mathbf{s}^{(k)}(a) = \left(1, s_2^{(k)}, s_3^{(k)}, ..., s_n^{(k)}\right) = \left(1, \left(d_2 \sigma_2^{(k)}\right), \left(d_3 \sigma_3^{(k)}\right), ..., \left(d_n \sigma_n^{(k)}\right)\right), k = 1, 2, .., q . \quad (23)$$

We remind that always $s_1 = +1$, and $d_l = \prod_{i=1}^{l-1} \varepsilon_{i,i+1}$, $l = 2, 3, ..., n$.

*Periodic boundary conditions.* Due to the possibility of the cyclic shift of the matrix elements (see the beginning of Subsection 4.1) without loss of generality we can assume that the value of $a_{1n}$ is the smallest in the absolute value. As above, Eq. (23) relates the max-configurations of the $F(a)$- and $F(m)$-functionals. If in the sequence $a_{12}, a_{23}, ..., a_{n-1,n}, a_{1n}$ the number of the negative coefficients is even, the spectrums of these two $F$-functionals coincide again. However, if the number of the negative coefficients is odd the different values of the spectrums of the $F(a)$- and $F(m)$-functionals differ from each other by $\pm 4 \cdot |a_{1n}|$: $F^{(k)}(a) = F^{(k)}(m) \pm 4 \cdot |a_{1n}|$, $k = 1, 2, ..., q$.

**2). Relation between spectrums and max-configurations of $F$-functionals for periodic and open boundary conditions.**

We showed how to define the max-configurations in the case of open boundary conditions (see Section 2, Theorem, and so on). The question is: can we define them when the boundary conditions are periodic: $a_{1n} \neq 0$? The results of computer simulations allow us to formulate the answer. Let $F(o)$ and $F(p)$ be the spectrums of the $F$-functionals in the cases of open and periodic boundary conditions, respectively. As before we suppose that the matrix element $a_{1n}$ is the smallest in the absolute value (see Eq. (17)). Then the sets of the max-configurations of the $F(p)$- and $F(o)$-functionals coincide. The values of the spectrums of these functionals differ by $\pm 2a_{1n}$:

$$F^{(k)}(p) = F^{(k)}(o) \pm 2 \cdot |a_{1n}|, \ k = 1, 2, ..., q .$$

So, in the general case of an arbitrary value of the element $a_{1n}$ the algorithm for finding the spectrum of the $F_p$-functional consists of three steps.

1. We cyclically shift $n$ coefficients $a_{i,i+1}$ by $k$ positions to the left so that the coefficient smallest by modulus occurs at the place of $a_{1n}$ and suppose that this element is equal 0.
2. For the obtained sequence of $n-1$ nonzero coefficients we make use of Theorem from Section 2 to find all max-configurations of the $F(o)$-functional.
3. We cyclically shift the coordinates of the obtained max-configurations at $k$ positions to the right and find the set of the max-configurations of the $F(p)$-functional for the case of periodic boundary conditions.

5. **Conclusions**

In this paper we develop and analyze a new method for finding all the $n$-dimensional configurations $\mathbf{s} = (s_1, s_2, ..., s_n)$, $s_i = \pm 1$ providing maxima of the quadratic functional $F(\mathbf{s}) = (\mathbf{A}\mathbf{s}, \mathbf{s})$ with the quasi-diagonal connection matrix:

$$\mathbf{A}_1 = \begin{pmatrix} 0 & a_{12} & 0 & 0 & \cdots & \cdots & a_{1n} \\ a_{21} & 0 & a_{23} & 0 & \cdots & \cdots & 0 \\ 0 & a_{32} & 0 & a_{34} & 0 & \cdots & 0 \\ 0 & \cdots & \cdots & \cdots & \cdots & \cdots & 0 \\ 0 & \cdots & \cdots & a_{n-2,n-3} & 0 & a_{n-2,n-1} & 0 \\ 0 & \cdots & \cdots & 0 & a_{n-1,n-2} & 0 & a_{n-1,n} \\ a_{n1} & 0 & \cdots & \cdots & 0 & a_{n,n-1} & 0 \end{pmatrix}. \qquad (24)$$

When we suppose that the matrix element $a_{1n} = 0$ we deal with the case of open boundary conditions. Then we have a chain of spins each of which interacts with its nearest neighbors. The inner local minima on the graph describing the dependence of the modulus of the matrix elements $\{|a_{i,i+1}|\}_1^{n-1}$ on the value of the index $i+1$ define the set of the max-configurations of the $F$-functional with such connection matrix (see Section 2). Introducing a nonzero interaction between the first and the last spins ($a_{1n} \ne 0$) we can roll the chain of spins into a ring and obtain the max-configurations for this system with the aid of the approach of Subsection 4.2.

We can generalize the aforesaid to the chain where each spins interacts not with the nearest but with the second neighbors. Then the connection matrix has the form:

$$\mathbf{A}_2 = \begin{pmatrix} 0 & 0 & a_{13} & 0 & 0 & \cdots & 0 & a_{1n-1} & 0 \\ 0 & 0 & 0 & a_{24} & 0 & 0 & \cdots & 0 & a_{2n} \\ a_{31} & 0 & 0 & 0 & a_{35} & 0 & \cdots & \cdots & 0 \\ 0 & a_{42} & 0 & 0 & 0 & a_{46} & \ddots & 0 & 0 \\ 0 & 0 & a_{53} & \ddots & 0 & \ddots & \ddots & 0 & 0 \\ \vdots & \vdots & \vdots & \ddots & \ddots & 0 & 0 & a_{n-3,n-1} & 0 \\ 0 & \vdots & \vdots & 0 & \ddots & 0 & 0 & 0 & a_{n-2,n} \\ a_{n-11} & 0 & \vdots & \cdots & 0 & a_{n-1,n-3} & 0 & \ddots & 0 \\ 0 & a_{n2} & 0 & \cdots & \cdots & 0 & a_{n,n-2} & 0 & 0 \end{pmatrix}. \qquad (25)$$

In Section 3 we presented the solution of the general problem when the only nonzero elements are on the second, third, ..., k-th, ... quasi-diagonals. To obtain these results we made use of the recursive algorithm allowing us to determine the optimal values of the spin coordinates (see Appendix).

A question to which we do not know the answer is as follows: can we construct the set of the max-configurations for the $F$-functional with the connection matrix $\mathbf{A}_{12} = \mathbf{A}_1 + \mathbf{A}_2$ if we know the sets of the max-configurations for the quasi-diagonal matrices $\mathbf{A}_1$ and $\mathbf{A}_2$ defined by Eq. (24) and Eq. (25)? We hope that if this is possible, we would be able to solve the problem of efficient maximization of the $F$-functional with an arbitrary connection matrix.

The other possibility is to develop for the matrix $\mathbf{A}_{12} = \mathbf{A}_1 + \mathbf{A}_2$ a new recursive algorithm similar to the one described in Appendix. We hope that in spite of the difficulty the problem has a solution. We plan to continue the investigation of this problem.

**Appendix**

In Appendix we present our proof of the theorem formulated in Section 2. We solve the system of equations (7) recursively. Namely, we choose one or another value of the initial spin and then the step by step use the requirement of the positiveness of the right-hand sides of the inequalities (7) to obtain the values of other spins $s_i$.

As previously, we base our reasoning on Fig. 1 where we show the dependence of the constants $\{a_{i,i+1}\}_{i=1}^{n-1}$ on the index $i+1$. On the intervals of the monotonic increase or decrease of the coefficients $\{a_{i,i+1}\}_1^{n-1}$ or when passing a local maximum in Fig. 1 we assign the value of the previous spin to the subsequent spin. However, when passing a point of the first inner local minimum in this figure, the sequence of the identical spins splits into two sub-sequences one of which consists of positive spins while in the other the spins are negative. Then at each new inner minimum on the graph of the coefficients $\{a_{i,i+1}\}_1^{n-1}$ each previously obtained sub-sequence splits into two new sub-sequences containing only positive or negative spins. The process continues till the end of the chain. For simplicity we assume all the coefficients $a_{i,i+1}$ to be positive.

**Proof of Theorem**. Let us analyze one after another all the inequalities in the right-hand sides of the set (7).

1). The first inequality (7.1) allows us to define the optimal values of two spin variables at once. The variable $s_1$ may be set equal +1 or -1. We take $\hat{s}_1 = +1$. Since all the constants $a_{i,i+1}$ are positive, from our supposition it follows that $s_2$ has also be positive: $\hat{s}_2 = +1$.

2). When $i = 2$, the inequality (7.2) takes the form $a_{21} + a_{23}s_3 > 0$. Since $a_{23} > a_{21}$ (see Fig. 1), the only possible value of the variable $s_3$ is $\hat{s}_3 = +1$. The aforesaid is true for every $i$ from the interval where the coefficients $a_{i,i+1}$ increase. Consequently, $\hat{s}_4 = \hat{s}_5 = ... = \hat{s}_{k+1} = +1$.

3). Inequality (7.2), which allows us to define the optimal value of the variable $s_{k+2}$, has the form

$$a_{k,k+1} + a_{k+1,k+2}s_{k+2} > 0.$$

Since now the coefficients $a_{i,i+1}$ decrease (see Fig. 1) this inequality has two solutions: $\hat{s}_{k+2} = +1$ and $\hat{s}_{k+2} = -1$. It turns out that the negative solution $\hat{s}_{k+2} = -1$ cannot be continued further. Indeed, if $\hat{s}_{k+2} = -1$, we have $h_{k+1}\hat{s}_{k+1} = a_{k,k+1} - a_{k+1,k+2} > 0$. However, at the next step the value of $h_{k+2}\hat{s}_{k+2} = -(a_{k+2,k+1} + a_{k+2,k+3}s_{k+3})$ remains negative for any value of the variable $s_{k+3}$. This means that from the two possible values $\hat{s}_{k+2} = \pm 1$ we have to leave only $\hat{s}_{k+2} = +1$.

Since at the interval $i+1 \in [k+3,l]$ the coefficients $a_{i,i+1}$ decrease steadily, we can use the same reasoning when determining the valid values of the variables $s_{k+3},...,s_l$. Thus, the optimal values of all these variables are +1: $\hat{s}_{k+3} = .... = \hat{s}_l = +1$.

4). Since $a_{l,l-1} > a_{l,l+1}$, the equation

$$h_l\hat{s}_l = a_{l,l-1} + a_{l,l+1}s_{l+1} > 0, \tag{A1}$$

for the optimal value of the variable $s_{l+1}$ has two solutions

$$\hat{s}_{l+1} = \pm 1. \tag{A2}$$

We can continue these both "seed" values to the larger indices (see below). Here the solution of the set of inequalities (7) splits into positive and negative branches. We will analyze each of these branches separately.

5). **Positive branch.** We set $\hat{s}_{l+1} = +1$ and take into account that in fact at the interval $i+1 \in [l+1,t+1]$ the dependence of the coefficients $a_{i,i+1}$ on the values of the indices is the same as at the previous interval $[2,l+1]$. Indeed, at first the coefficients $a_{i,i+1}$ increase and then decrease. Since the seed value $\hat{s}_{l+1}$ is positive, for other spin variables we obtain $\hat{s}_{l+2} = .... = \hat{s}_t = +1$. For the variable $s_{t+1}$ both values are possible: $\hat{s}_{t+1} = \pm 1$.

6). **Negative branch.** Let us examine what happens when the "seed" value in Eq. (A2) is negative:

$$\hat{s}_{l+1} = -1.$$

In this case Eq. (A1) has the form

$$h_l \hat{s}_l = a_{l,l-1} - a_{l,l+1} > 0$$

and for determination of the next variable $s_{l+2}$ we obtain the equality:

$$h_{l+1}\hat{s}_{l+1} = a_{l+1,l}\hat{s}_l\hat{s}_{l+1} + a_{l+1,l+2}s_{l+2}\hat{s}_{l+1} = -(a_{l+1,l} + a_{l+1,l+2}s_{l+2}) > 0. \tag{A3}$$

Since $a_{l+1,l} < a_{l+1,l+2}$, we can satisfy the inequality (A3) only if $\hat{s}_{l+2} = -1$. Continuing the same reasoning we obtain a sequence of negative values of spins:

$$\hat{s}_{l+2} = \hat{s}_{l+3} = \hat{s}_{l+4} = ... = \hat{s}_m = -1.$$

Taking into account that the right-hand side of the equation

$$h_m\hat{s}_m = a_{m,m-1}\hat{s}_{m-1}\hat{s}_m + a_{m,m+1}s_{m+1}\hat{s}_m = a_{m,m-1} - a_{m,m+1}s_{m+1}$$

is positive only if $\hat{s}_{m+1} = -1$, the variable $s_{m+1}$ may be only negative.

To define the value of the next variable $\hat{s}_{m+2}$ we have the chain of equalities

$$h_{m+1}s_{m+1} = a_{m+1,m}\hat{s}_m\hat{s}_{m+1} + a_{m+1,m+2}s_{m+2}\hat{s}_{m+1} = a_{m+1,m}(-1)(-1) + a_{m+1,m+2}s_{m+2}(-1) = a_{m+1,m} - a_{m+1,m+2}s_{m+2}.$$

The value of $a_{m+1,m} - a_{m+1,m+2}s_{m+2}$ is positive for any $\hat{s}_{m+2} = \pm 1$. However, from the next chain of equations

$$h_{m+2}\hat{s}_{m+2} = a_{m+2,m+1}\hat{s}_{m+1}\hat{s}_{m+2} + a_{m+2,m+3}s_{m+3}\hat{s}_{m+2} = a_{m+2,m+1}(-1)(+1) + a_{m+2,m+3}s_{m+3}(+1) = -a_{m+2,m+1} + a_{m+2,m+3}s_{m+3}$$ we obtain that the expression $-a_{m+2,m+1} + a_{m+2,m+3}s_{m+3}$ cannot be positive for any value of the variable $s_{m+3}$. Consequently, the we are left with the only possibility $\hat{s}_{m+2} = -1$.

It is easy to see that the next spin variables from $\hat{s}_{m+1}$ to $\hat{s}_t$ can have only negative values:

$$\hat{s}_{m+1} = \hat{s}_{m+2} = \hat{s}_{m+3} = ... = \hat{s}_t = -1.$$

Thus, we found that when the sequence of the negative solutions of the set of inequalities (7) passes through the point of the maximum of the coefficients the sequence does not split into two subsequences: it behaves the same as the sequence of the positive solutions at the point where the values of the coefficients achieve their maximum.

It is easy to verify that the sequence of the negative solutions splits at the point $t+1$: $\hat{s}_{t+1} = \pm 1$. Each of the two obtained branches can be continued by the spins with the same signs:

$$\hat{s}_{t+1}^{(-)} = \hat{s}_{t+2}^{(-)} = ... = -1 \text{ and } \hat{s}_{t+1}^{(+)} = \hat{s}_{t+2}^{(+)} = ... = +1.$$

**7).** With the aid of the graph in Fig. 1 with three inner minima we described in detail how the splitting of the spin sequences occur. Let we have $p$ inner minima. Then at the minimum points each of the current spin sequence splits into two branches of positive and negative spins. When the number of the inner minima equals $p$, the whole number of the sequences equals $2^p$. Since at each step we guarantee fulfillment the inequality $h_i s_i > 0$ these sequences provide maxima of the $F$-functional. When the index $i \in [1, n]$, there are no other sequences for which this inequality can be satisfied.

**8).** Here we explain how one can calculate the contribution of the splitting points to the value of the $F$-functional. When a sequence of positive spins passes a splitting point $\hat{s}_{r+1} = \pm 1$ the contributions to the $F$-functional differ depending on the positive or negative branch the sequence goes. We denote these two possibilities as $(+) \to (+)$ and $(+) \to (-)$, respectively. The most important are the values of the sums $\hat{h}_r\hat{s}_r + \hat{h}_{r+1}\hat{s}_{r+1}$ on the different branches:

$$\hat{h}_r \hat{s}_r + \hat{h}_{r+1} \hat{s}_{r+1} = \begin{cases} a_{r-1,r} + 2a_{r,r+1} + a_{r+1,r+2}, & (+) \to (+) \\ a_{r-1,r} - 2a_{r,r+1} + a_{r+1,r+2}, & (+) \to (-). \end{cases} \quad (A4)$$

In the same way, when a sequence of negative spins passes the splitting point $\hat{s}_{r+1} = \pm 1$, the term $\hat{h}_r \hat{s}_r + \hat{h}_{r+1} \hat{s}_{r+1}$ differs depending on which branch – positive or negative – the transition occurs:

$$\hat{h}_r \hat{s}_r + \hat{h}_{r+1} \hat{s}_{r+1} = \begin{cases} a_{r-1,r} + 2a_{r,r+1} + a_{r+1,r+2}, & (-) \to (-) \\ a_{r-1,r} - 2a_{r,r+1} + a_{r+1,r+2}, & (-) \to (+). \end{cases} \quad (A5)$$

Equations (A4) and (A5) are necessary when calculating the values of the local maxima of the $F$-functional.

## Acknowledgements

I am grateful to my long-term colleagues Boris Kryzhanovsky and Yakov Karandashev from discussions with whom a lot ideas for this work were born.

Remarks of Inna Kaganova who translated this paper were helpful.